\documentclass[conference]{IEEEtran}
\IEEEoverridecommandlockouts
\usepackage{cite}
\usepackage{amsmath,amssymb,amsfonts}
\usepackage{algorithmic}
\usepackage{graphicx}
\usepackage{textcomp}
\usepackage{xcolor}
\usepackage{dblfloatfix}
\usepackage{booktabs}
\usepackage{tabularx}
\usepackage{tikz}
\usepackage[T1]{fontenc}
\usepackage[utf8]{inputenc}
\usetikzlibrary{shapes,arrows}
\def\BibTeX{{\rm B\kern-.05em{\sc i\kern-.025em b}\kern-.08em
    T\kern-.1667em\lower.7ex\hbox{E}\kern-.125emX}}

\tikzstyle{block} = [draw, fill=purple!16, rectangle, 
    minimum height=3em, minimum width=6em]
\tikzstyle{input} = [coordinate]
\tikzstyle{output} = [coordinate]
\tikzstyle{pinstyle} = [pin edge={to-,thin,black}]

\usepackage[bottom]{footmisc}

\makeatletter
\def\ps@IEEEtitlepagestyle{%
  \def\@oddfoot{\mycopyrightnotice}%
  \def\@evenfoot{}%
}
\def\mycopyrightnotice{%
  \begin{minipage}{\textwidth}
  \footnotesize
  978-1-XXXX-XXXX-X/XX/0 \copyright 2023 IEEE \hfill
  \end{minipage}%
  \gdef\mycopyrightnotice{}
}
\makeatother

\begin{document}
\title{Vision Transformer for Classification of UAV and Helicopters Using Micro-Doppler Spectrograms in Surveillance Radar\\}

\author{\IEEEauthorblockN{Arkadiusz Czuba}
\IEEEauthorblockA{\textit{Warsaw University of Technology} \\
\textit{PIT-RADWAR S.A.}\\
Warsaw, Poland \\
arkadiusz.czuba.dokt@pw.edu.pl}

}

\maketitle
\begin{abstract}
\\Machine learning researchers strive to develop better and better algorithms to solve computer vision problems, such as image classification. In recent years, the classification of micro-Doppler spectrograms has also benefited from these findings. Convolutional neural networks (CNNs) became the gold standard for these tasks. Unfortunately, CNNs can work on fixed-resolution images, or they need to resize mismatched images to fit input dimensions. It can become a problem when micro-Doppler spectrograms are generated with e.g. different integration times. The goal of this work was to classify the UAV and helicopters micro-Doppler spectrograms with different duration times, using the Vision Transformer (ViT) architecture. Before that, spectrograms signal-to-noise-ratio and micro-Doppler features visibility were improved by denoising algorithm based on modified Dual Tree Complex Wavelet Transform. The experiments were conducted on real data collected using surveillance, short range, military radar. As a result, it has been shown that the ViT model achieved 97.76\% accuracy for this task. To further interpret the network performance, the raw self-attention maps were analyzed.
\\
\end{abstract}


\begin{figure*}[b]
  \centering
  \includegraphics[width=14cm,height=4cm]{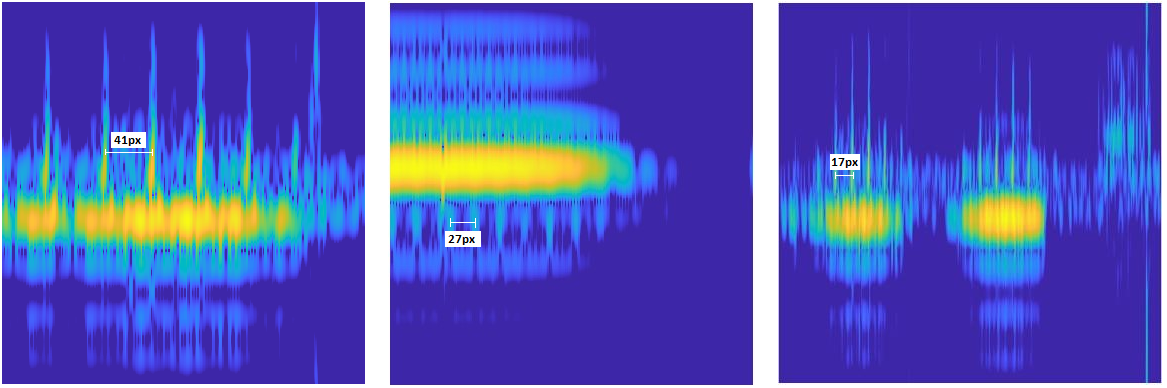}
  \caption{Changing the size of images can distort the micro-Doppler signature (vertical axis represents frequency, horizontal axis represents time). }
  \label{fig:cnn_resize}
\end{figure*}

\section{Introduction}

The role of rotorcrafts in the modern battlefield has gained more importance in recent years. The ever-growing number of Unmanned Aerial Vehicles (UAVs) and their small size poses a challenge to radar systems. 

The micro-Doppler effect, introduced by Chen \cite{chen_micro-doppler_2006}, is the Doppler scattering signal modulation produced by micro-motions of mechanical vibration or rotation of the target elements. The features of the detected target can be extracted from the analysis of the micro-Doppler signal return. Many different micro-Doppler signal representation methods are studied \cite{gerard_micro-doppler_2021}, but the most common is the spectrogram obtained from the short-time Fourier transform (STFT). 

Unfortunately, the radar dwell time has to be long enough to be suitable for micro-Doppler signal detection and classification. In surveillance radars, the faster antenna rotation rate typically translates to a shorter dwell time. These challenging conditions may result with often insufficient temporal resolution \cite{gong_detection_2022}, \cite{ma_classification_2021}. Moreover, the signal-to-noise ratio (SNR) has to be high enough for accurate classification \cite{dale_snr-dependent_2022}. The micro-Doppler classification can achieve high accuracy when the SNR is high and the radar dwell time is long. In operational systems, that is generally not the case. Moreover, surveillance radar systems can operate in different modes e.g., longer/shorter range, rotating or stationary antenna etc. The coherent processing interval (CPI) duration can vary in different radar operation modes. Also, various classes of targets can require different integration times to achieve good micro-Doppler signatures. Assuming that SNR is high, the faster the changes in micro-Doppler oscillations, the shorter the integration time can be. On the other hand, the slower changes, the more extended integration is mandatory.

The recent advances in deep learning \cite{lecun_deep_2015} have changed machine learning research by leaning towards deep neural networks instead of hand-crafted feature extraction-based classifiers. In radar systems, neural networks are widely used in many applications, such as automatic target recognition, classification, jamming and clutter recognition, interference suppression, and array design \cite{geng_deep-learning_2021}, \cite{lang_comprehensive_2020}. Deep learning is also one of the technologies that support the development of cognitive radar \cite{smith_neural_2020}. In studies on micro-Doppler spectrograms classification, the primary attention gained convolutional neural networks (CNNs), adopted from computer vision. In \cite{huizing_deep_2019}, authors achieved excellent performance of mini-UAVs spectrograms classification using classical CNN configuration. Classical CNN architecture can also achieve high accuracy in low-SNR classification scenarios \cite{raval_convolutional_2021}. Moreover, the use of pre-trained weights and more complex architectures, such as GoogLeNet \cite{szegedy_going_2015} or ResNet\cite{he_deep_2016}, can lead to solving more demanding problems with larger datasets, and with more accuracy \cite{gerard_micro-doppler_2021}, \cite{dale_snr-dependent_2022}, \cite{rahman_classification_2020}, \cite{kim_drone_2017}. The need for more extensive and real scenario datasets was mentioned in detail here \cite{rahman_classification_2020}.

Apart from convolution-based models, the Transformer models are gaining more momentum since they were first proposed in \cite{vaswani_attention_2017}. Transformer was first used in natural language processing applications \cite{devlin_bert_2019} and later found its place in computer vision \cite{dosovitskiy_image_2021}. Vision Transformer (ViT) is a competitor to convolution-based models. One of the advantages of Transformer-based networks over CNNs is the ability to represent images as sequences of patches of varying lengths. The CNNs learn data structures with fixed input dimensions. CNN input image has to be resized to the required input layer dimensions. Unfortunately, resizing the spectrograms can degrade the micro-Doppler signatures by changing the actual distances between features. The example of resizing the micro-Doppler spectrogram is shown in Fig.~\ref{fig:cnn_resize}. Left and middle images represent the 102.4 ms spectrogram of the helicopter micro-Doppler signature and drone, respectively. The image on the right shows resized 307.2 ms spectrogram of the same helicopter to fit the resolution of the previous ones. From this micro-Doppler signature analysis, it can be deduced that the helicopter blades are spinning faster than the UAV. This fact can mislead the classifier, especially when the number of classes is extensive, and their differences become subtle.

The goal of this study is to classify the micro-Doppler spectrograms with various duration time. To complete this task, ViT architecture was chosen. The real data of the UAV and helicopters were collected by military surveillance radar with a mechanically rotating antenna.

The remainder of the paper is organized as follows. Section II introduces the essential concepts of surveillance radar operation and its parameters. Moreover, the dataset used in this work, the signal processing that led to its generation, and the spectrogram denoising algorithm are discussed. Section III focuses in detail on the ViT model, its building components, training, and hyperparameters. Section IV presents the experimental results of the classification system introduced in this work and analyses the achieved results.

\begin{figure*}[t]
  \centering
  \includegraphics[width=15cm,height=8cm]{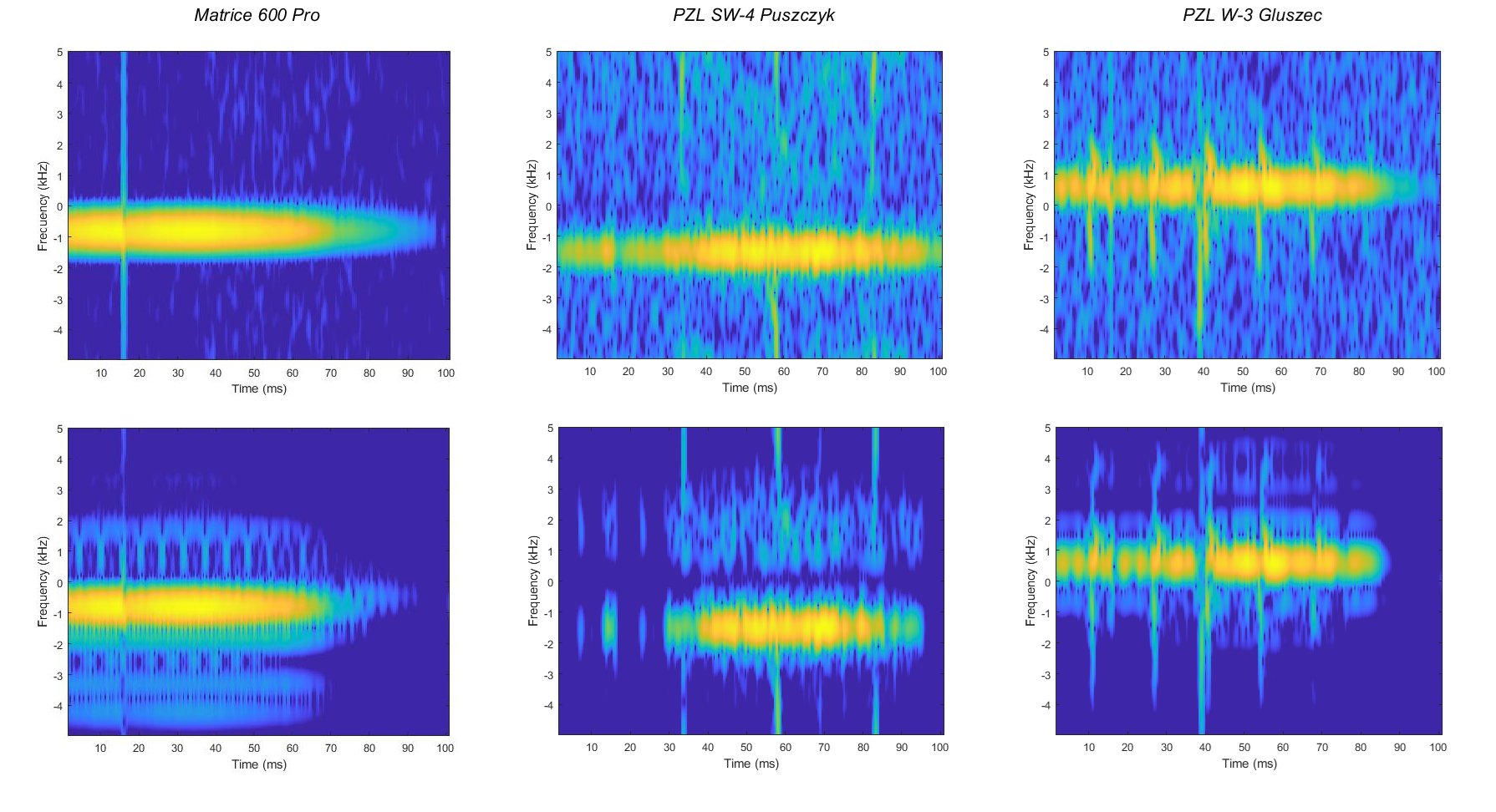}
  \caption{Micro-Doppler spectrograms of a drone Matrice 600 Pro, and helicopters PZL SW-4 Puszczyk and PZL W-3 Gluszec. Each column of this figure contains spectrograms before (upper row) and after (lower row) denoising with modified DTCWT denoising algorithm. }
  \label{fig:spectr}
\end{figure*}

\section{Micro-Doppler Spectrograms}

\subsection{Surveliance Radar and Targets}

The data was collected using the Sola redeployable radar system \cite{noauthor_solredeployable_nodate}. It is a multi-mission three-dimensional surveillance radar manufactured by PIT-RADWAR. Besides typical aircrafts, Sola radar can detect mortar bombs, hoovering helicopters, and UAVs with low radar cross section (RCS). Mechanically rotated active antenna provides 360 degrees of azimuth coverage and up to $\sim$55 degrees in elevation. The antenna rotation speed during data collection was 30 rpm. A separate antenna channel for rotorcraft target detection operates in the C band. It works in a high pulse repetition frequency (PRF) mode that provides high temporal resolution. To increase the dwell time beam has $\sim$10 degrees width for -3dB level.

Data was collected during the data collection campaign. Three types of targets were illuminated: UAV Matrice 600 Pro, PZL SW-4 Puszczyk, a light single-engine helicopter, and multipurpose battlefield helicopter PZL W-3 Gluszec. Rotorcrafts are consisting six rotors with two blades each, a three-blade rotor and a four-blade rotor, respectively.

\subsection{Spectrograms Generation and Denoising}

The collected radar raw data consist of complex-valued samples after quadrature demodulation. To obtain the micro-Doppler spectrogram, the signal processing channel firstly detects and recognizes the features of the rotorcraft \cite{misiurewicz_analysis_1997}, \cite{misiurewicz_experiments_2002}, such as a hub or rotor blades. Only CPIs with detected rotor blade flashes were considered for classification in this work. The 1024, 2048, and 3072 pulses were taken for spectrogram generation. The CPIs originate from only one range bin. The design of radar signal processing ensures that at least one blade flash will be visible in every micro-Doppler spectrogram. It is worth mentioning that a highpass filter was applied to cut off the near-zero Doppler frequency components before further processing, meaning echoes of stationary objects and ground clutter.

For PRF equal to 10kHz, the spectrogram time is 102.4 ms. The spectrograms were generated using the STFT with the Blackman-Harris window. The window size was set to 32 samples with 80\% overlap between them. The spectrogram examples of raw data are presented in the first row of Fig.~\ref{fig:spectr}.

The denoising algorithm based on Dual-Tree Complex Wavelet Transform (DTCWT) was applied to increase the SNR. Since we are dealing with complex (in-phase and quadrature - IQ) signals, typical DTCWT could not be used. Denoising the IQ components as two separate channels is a method of last resort that should be avoided. Instead, the denoising algorithm was based on modified DTCWT (mDTCWT) \cite{serbes_modified_2011} dedicated to quadrature signals. The best results were achieved by using the soft-thresholding function and with 3-level signal decomposition. The results are shown in the second row of Fig.~\ref{fig:spectr}. Denoising of Matrice 600 Pro spectrograms revealed blade flashes previously entirely covered by background noise. Overall, the algorithm improved the SNR and the visibility of micro-Doppler features. Denoising SNR and peak SNR (PSNR) results are presented in Table.~\ref{denoising-tab}.

A block diagram of the classification pipeline used in this work is shown in Fig.~\ref{fig:block_dig}. A highpass filter filters \textit{IQ} samples. Output from the filter \textit{h} is denoised by the mDTCWT-based algorithm and transformed by STFT to obtain the spectrogram image \textit{img}. Finally, the neural network output provides information about the \textit{class} of the detected target.

\subsection{Dataset}

The collected dataset was split into 80\% data for training, 10\% for validation, and 10\% for testing. The dataset contains a total of 13614 spectrograms, with the data evenly distributed among classes except for PZL SW-4 Puszczyk, which is slightly more numerous than the others. Further, the dataset is split evenly into three parts, where each contains spectrograms with a duration time equal to 102.4, 204.8, or 307.2 ms. The spectrogram images for each duration time have 384x384, 384x768, and 384x1152 resolution, respectively, with three RGB color channels.

\section{Vision Transformer for Micro-Doppler Classification}

\subsection{Neural Network Model}\label{vit}

\begin{figure*}[b]
\centering
    % The block diagram code is probably more verbose than necessary
    \begin{tikzpicture}[auto, node distance=2cm,>=latex']
        % We start by placing the blocks
        \node [input, name=input, node distance=1cm] {};
        % \node [sum, right of=input] (sum) {};
        \node [block, right of=input] (highpass) {Highpass filter};
        \node [block, right of=highpass ,node distance=3.2cm] (controller) {mDTCWT};
        \node [block, right of=controller, node distance=3.2cm] (system) {STFT};
        \node [block, right of=system, node distance=3.2cm] (model) {Neural Network};
        % We draw an edge between the controller and system block to 
        % calculate the coordinate u. We need it to place the measurement block. 
        \draw [->] (controller) -- node[name=u] {$d$} (system);
        \draw [->] (system) -- node {$img$} (model);
        \node [output, right of=model, node distance=2.4cm] (output) {};
        
        % Once the nodes are placed, connecting them is easy. 
        \draw [draw,->] (input) -- node {$IQ$} (highpass);
        \draw [->] (highpass) -- node {$h$} (controller);
        
        \draw [->] (model) -- node [name=y] {$class$}(output);
        
    \end{tikzpicture}
    \caption{Block diagram of a complete classification system.}
   \label{fig:block_dig}
\end{figure*}
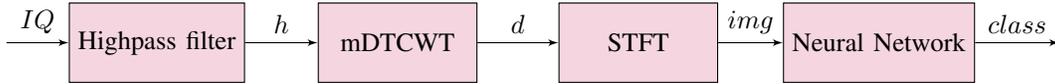

The configuration and training guidelines of the ViT model used in this work are based on the ViT-S model mentioned in \cite{steiner_how_2022}. The modified ViT-S (ViT-S-M) structure is shown in Fig.~\ref{fig:block_dig}. The input spectrogram image is split into a series of 32x32 patches, flattened, and transformed into a patch embedding feature map by a single convolutional layer. Spectrograms of length 102.4 ms and 204.8 ms were padded to fit 384x1152 resolution. The padding mechanism was inspired by the BERT architecture \cite{devlin_bert_2019}. The sequence of the 2D embeddings is then concatenated with position embeddings and learnable class token. The length p of patch embeddings depends on the image resolution. In this work the \textit{p} is equal to 433. The main layer of ViT architecture is the Transformer encoder (TE). The original ViT-S architecture, composed of 12 TE layers, is too large for this classification task and caused significant overfitting. Instead, only 5 TE layers were used. The single TE layer consists of multi-head self-attention (MSA) and multilayer perceptron (MLP) blocks. The LayerNorm precedes each block, and the residual connection succeeds each block. The classifier is implemented as a single MLP layer with LayerNorm. 

\subsection{Self-attention Maps}\label{map}

\begin{table}[b]
  \caption{Spectrogram denoising results for each target class}
  \label{denoising-tab}
  \centering
  \begin{tabular}{l c c}
    \toprule
    Target class            &        SNR (dB)       &       PSNR (dB)     \\
    \midrule
    Matrice 600         &       11.96      &	  17.63  \\
    PZL SW-4 Puszczyk   &       9.02	    &     13.12  \\
    PZL W-3 Gluszec     &       10.83   	&     15.35  \\
    \bottomrule
  \end{tabular}
\end{table}

The MSA block with a self-attention mechanism is the most crucial part of any Transformer network. Self-attention is used to find pairwise interactions between input vectors to learn hierarchies in input data. In essence, the attention mechanism relates different sequence positions to search for a representation of that sequence \cite{vaswani_attention_2017}. The self-attention mechanism can be visualized as the raw attention map \cite{chefer_transformer_2021}. 

\begin{figure*}[b]
  \centering
  \includegraphics[width=10cm,height=7cm]{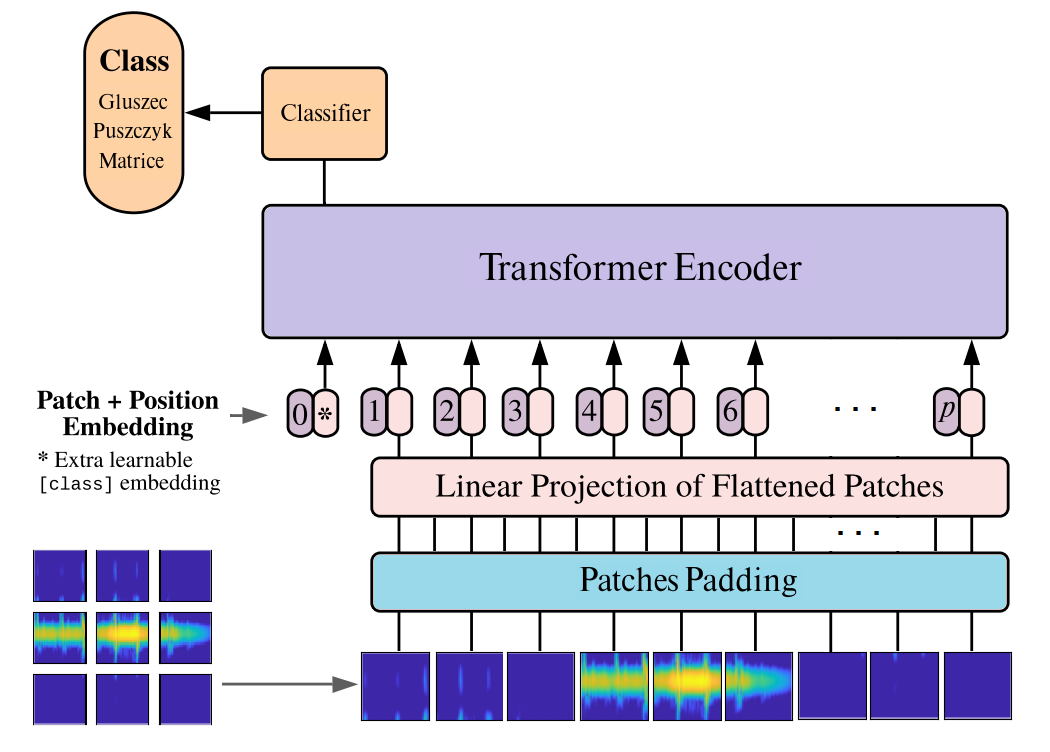}
  \caption{Block diagram of a complete classification system (inspired by \cite{dosovitskiy_image_2021}).}
  \label{fig:network_block}
\end{figure*}

\subsection{Hyperparameteres and Training}

The fine-tuning of neural network hyperparameters is critical to achieving the best model performance. Like the ViT configuration from Section III-A, the training parameters and hyperparameters are strongly inspired by the paper \cite{steiner_how_2022}. 

The network was trained with Adam optimizer \cite{kingma_adam_2017} with $\beta= 0.9$ and $\beta= 0.999$. Instead of cosine learning rate scheduling with a warm-up, the stepped scheduling decays the learning rate by $\gamma=0.5$ every 12 epochs. The initial learning rate was set to $2*10^{-3}$. The batch size set for this problem was 32. As we deal with a multi-class classification problem, the cross-entropy loss function has been chosen. Unfortunately, the model had susceptibility to overfitting the training data. After numerous experiments, the best results were achieved with L2 regularization $\lambda=10^{-4}$, and with 10\% dropout after MLP and embedding layer. There is no dropout after MSA layers. The model training process was set to 80 epochs. The neural network training takes no more than eight hours on NVIDIA GeForce GTX 1060 GPU with 1.582 GHz clock frequency.

\begin{table*}[!b]
  \caption{The comparison of the ViT-Ti and GoogLeNet technical parameters.}
  \label{vit-params}
  \centering
  \begin{tabular}{l c c c c r}
    \toprule
    Model     &      Image resolution  &       No. of layers & No. of learnable parameters (M)   & Model size (GB) & MAC (G) \\
                                                                     
    \midrule
    GoogLeNet \cite{szegedy_going_2015}  &  224x224	&   22	&    10.9    &	1.6	    &    82   \\            
    GoogLeNet 	     &  384x384	     &   22	     &    10.9    &	4.6	    &    242     \\
    GoogLeNet 	     &  384x1152	&   22	     &    10.9    &	13.8	    &    728    \\
    GoogLeNet 	     &  384x1536	&   22	     &    10.9    &	18.5	    &    971    \\
    ViT-S-M 	     &  384x1152	&   5 (TE)	&    10.6    &	1.7	    &    16.5   \\
    ViT-S-M  	     &  384x1536	&   5 (TE)	&    10.6    &	2.2	    &    21.9  \\
    \bottomrule
  \end{tabular}
\end{table*}

% \begin{figure*}[b]
% \centering
% \begin{minipage}{.7\textwidth}
%   \centering
%   \includegraphics[width=1.1\linewidth]{loss_final_rev.png}
%   \caption{Vision Transformer training loss and accuracy characteristics}
%   \label{fig:loss}
% \end{minipage}%
% \begin{minipage}{.5\textwidth}
%   \centering
%   \includegraphics[width=.6\linewidth]{conf_mat_rev.png}
%   \caption{Vision Transformer classification results}
%   \label{fig:conf-mat}
% \end{minipage}
% \end{figure*}

\section{Experimental Results}

\begin{figure*}[!b]
  \centering
  \includegraphics[width=14cm,height=6.5cm]{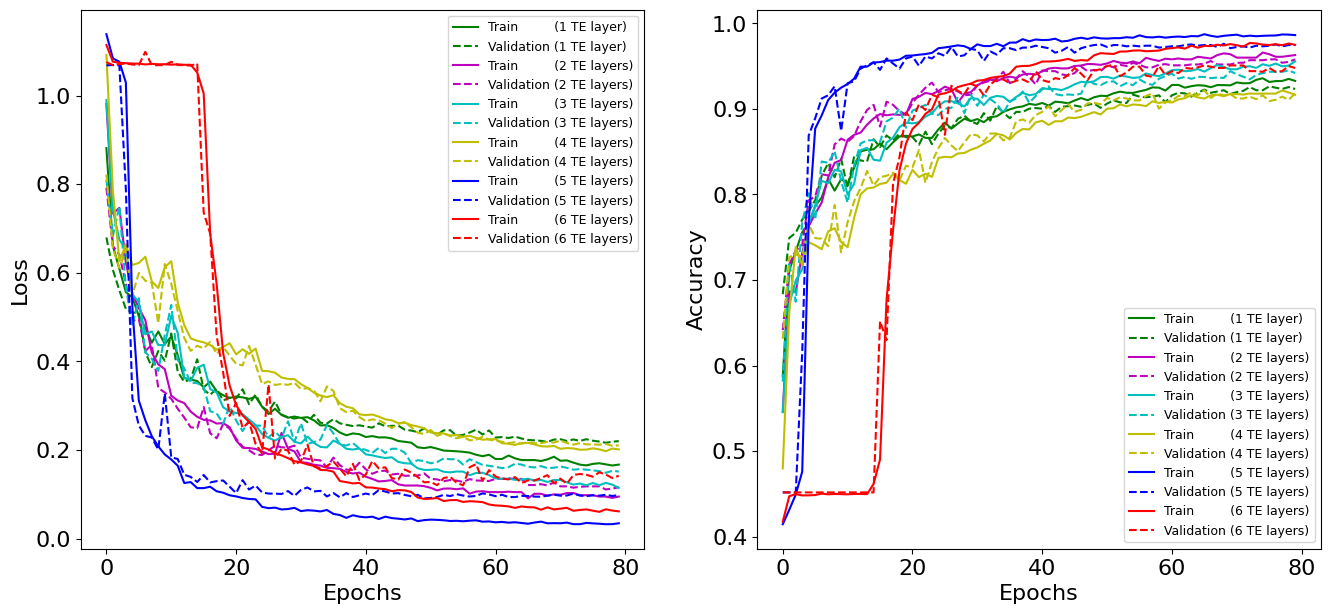}
  \caption{Vision Transformer training loss and accuracy characteristics for different number of TE layers.}
  \label{fig:loss}
\end{figure*}

\begin{figure*}[!t]
  \centering
  \includegraphics[width=15cm,height=8cm]{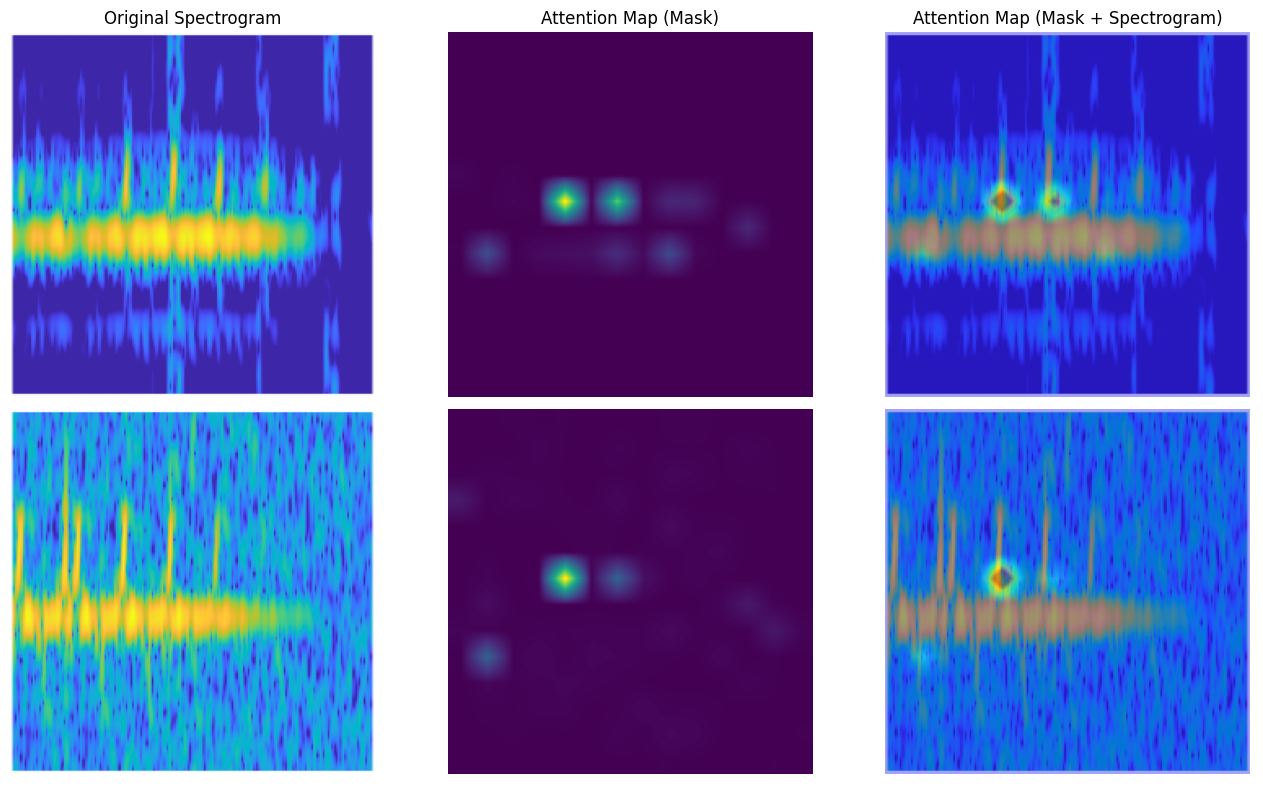}
  \caption{Examples of the raw attention maps of PZL W-3 Gluszec helicopter spectrograms after denoising (upper row) and before denoising (lower row).}
  \label{fig:att}
\end{figure*}

In this section, the experimental results of the micro-Doppler classification with ViT-S-M are presented. 

Initial training with the original VIT-S configuration composed of 12 TE layers caused significant overfitting due to excess learnable parameters. Numerous experiments were conducted to find the optimal number of TE layers for this task. The results of training the ViT-S-M with different numbers of TE layers are shown in Fig.~\ref{fig:loss}. The training process with 1,2,3 and 4 TE is quite successful, but the classification performance is not impressive. It's probably caused by an insufficient number of learnable parameters. The ViT-S-M with 5 TE reached excellent performance and still avoided overfitting. After that, the network trained with 6 TE accuracy dropped, and training loss started to outperform validation loss, meaning the ViT-S model started to overfit. Overall, ViT-S-M network training and validation accuracy achieved 98.73\% and 97.61\%, respectively. The classification performance of the ViT-S-M on the test set reached an accuracy of 97.76\%. 

The network technical parameters were confronted with state-of-the-art GoogLeNet neural network architecture. GoogLeNet, as mentioned in Section I, was previously used for micro-Doppler spectrogram classification. Let us assume that we can also pad and crop the images to fit the CNN input without degradation of the micro-Doppler signature but bearing the cost of larger input layer dimensions. The results of that comparison are presented in Table.~\ref{vit-params}. Despite the depth difference, both models have almost the same number of learnable parameters. Firstly, the GoogLeNet model, with an increase in input image resolution, drastically grew in size and multiply-accumulation (MAC) operations. Eight times more memory and close to forty-four times more computational power are required to design the equivalent of ViT-S-M as a GoogLeNet model with 384x1152 input resolution. Secondly, the ViT-S-M model with an original resolution of 384x1152 has almost the same size as GoogLeNet with a 224x224 input resolution and still requires five times fewer MAC operations to calculate. The ViT model with 384x1536 input resolution shows that the increase in patch embedding length in ViT slightly increases the model size and required memory space. The technical parameters analysis shows that the ViT-S-M requires much less computational power. That information is crucial when choosing a hardware platform for model deployment. The Xilinx Versal ACAP platform \cite{noauthor_versal_nodate} fulfills radar systems' low latency and high-throughput military requirements.

The examples of micro-Doppler raw attention maps for denoised and original spectrograms are shown in  Fig.~\ref{fig:att}. The green and red spots on the \textit{Mask} and \textit{Mask + Spectrograms} figures represent the high attention scores. For the input spectrogram with high SNR, the MSA has found three blade flashes with different intensities. On the other hand, the spectrogram with lower SNR has less accurate attention scores, but it can be observed that MSA partially found micro-Doppler features despite the noise.

\section{Conclusion}

The goal of this work was to classify micro-Doppler spectrograms with various lengths with the Vision Transformer architecture. The spectrograms were denoised using a modified DTCWT to improve the SNR of the micro-Doppler features and reveal those covered by background noise. Firstly, it has been shown that micro-Doppler spectrograms generated with various integration times can be classified with success. Moreover, the Vision Transformer neural network was successfully trained and evaluated. The classification accuracy of 97.76\% was achieved. Compared to GoogLeNet architecture, ViT model shows has much better resource utilization and requires less computational power for the same task. Also, the analysis of raw self-attention maps revealed the fairly good noise robustness of the attention mechanism. However, the presented work is far from the exhaustion of the topic. Future work includes expanding the problem to object detection, especially in longer spectrograms that may contain more than one target. The author's priority is also integrating this work in the cognitive radar system.

\bibliographystyle{IEEEtran}
\bibliography{references.bib}

\end{document}